\documentstyle[psfig,prb,aps]{revtex}
\begin{document}
\draft

\twocolumn[\hsize\textwidth\columnwidth\hsize\csname
@twocolumnfalse\endcsname

\title{
Lattice distortions around a Tl$^+$ impurity in NaI:Tl$^+$ and CsI:Tl$^+$
scintillators. An {\em ab initio} study involving large active clusters.
}
\author{Andr\'es Aguado}
\address{Departamento de F\'\i sica Te\'orica, Facultad de Ciencias,
Universidad de Valladolid, 47011 Valladolid, Spain}
\author{Andr\'es Ayuela}
\address{Laboratory of Physics, Helsinki University of Technology,
02015 Espoo, Finland}
\author{Jos\'e M. L\'opez and Julio A. Alonso}
\address{Departamento de F\'\i sica Te\'orica, Facultad de Ciencias,
Universidad de Valladolid, 47011 Valladolid, Spain}
\maketitle
\begin{abstract}
{\em Ab initio} Perturbed Ion cluster-in-the-lattice
calculations of the impurity centers NaI:Tl$^+$
and CsI:Tl$^+$ are presented. We study several active clusters of increasing
complexity and show that the lattice relaxation around the Tl$^+$ impurity
implies the concerted movement of several shells of neighbors.
The results also reveal the importance of considering a set of ions that can
respond to the geometrical displacements of the inner shells by adapting
selfconsistently their wave functions. Comparison with other
calculations involving comparatively small active clusters
serves to assert the significance of our conclusions. 
Contact with experiment
is made by calculating absorption energies. These are in excellent
agreement with the experimental data for the most realistic active clusters
considered.
\end{abstract}

\vskip2pc]

\section{Introduction}
\label{aguado:intro}

The interest in luminescent materials has increased in
recent years because of their numerous technological applications. \cite{Bla95}
Most of these materials involve the doping of a pure ionic crystal, that is 
substitution of some of the ions by other ions with specific
absorption-emission characteristics (Tl$^+$, Ce$^{3+}$, etc). The presence of
impurity ions in ionic crystals induces geometrical distortions in the host
lattice. Those distortions are an important ingredient in the scintillation
process and thus it is important to establish a reliable description
of them.

A convenient way to calculate the lattice distortions is to model the doped
crystal by a finite cluster centered on the impurity and embedded in a
field representing the rest of the host lattice. This cluster approach
has been used to study the geometrical and optical properties
of doped crystals. \cite{Che81,Win87,Kun88,Bar88,Lua89a,Lua89b,Sei91,And91,Isl92,Lua92,Miy93,Sca93,Lua93,Vis93,And93,Flo94,Bel94,Ber95,Pas95,Gry95,Llu96,Sei96,Ber96a,Ber96}
The cluster (active space) can be studied by using standard
quantum-mechanical methods. The rest of the crystal (environment) can be 
described in several ways. In the simplest and most frequently used approach,
the environment is simulated by placing point charges on the
lattice sites, but
this procedure has to be improved in order to obtain a
realistic description of the lattice distortions around the impurity.
\cite{Bar88,Lua89a,Lua89b,Sei91,Lua92,Lua93,Flo94,Pas95,Llu96,Sei96}
Model potentials have been developed to represent the effects of the
environment on the active cluster, that include attractive and repulsive
quantum-mechanical
terms aside from the classical Madelung term, \cite{Huz87} but a problem
still remains: the large computational cost of conventional molecular orbital
(MO) calculations prevents from performing an exhaustive geometrical relaxation of the
lattice around the impurity. In the most accurate MO calculations,
\cite{Sei91,Pas95,Llu96,Sei96,Ber96} only the positions of the ions in the
first shell around the impurity are
allowed to relax. However,
geometrical relaxations far beyond the first shell of
neighbors can be expected.
In fact, recent classical simulations of solids, 
\cite{Zha93,Zha94,Isl94,Isl95,Say95,Akh95,Exn95,Cat98a,Cat98b} 
performed employing
phenomenological potentials, \cite{Sto81,Har90,Wil93,Wil96,Gal96,Gal97}
have shown the importance of considering
appropriate large-scale lattice relaxations in the study of a variety of
intrinsic and extrinsic defects in ionic crystals.

In this contribution we report theoretical calculations of the
lattice distortions induced by a Tl$^+$ impurity in two well-known ionic
scintillators, namely NaI:Tl$^+$ and CsI:Tl$^+$. For this purpose we use the
{\em ab initio} Perturbed Ion (PI) model,
\cite{Lua90a,Lua90b,Pue92,Lua92a,Lua93a} which circunvents the
problems mentioned above: (a) The active cluster is embedded in an environment
represented by the {\em ab initio} model potentials of Huzinaga 
{\em et al.} \cite{Huz87}; (b) The computational simplicity of the PI model
allows for the geometrical relaxation of several coordination shells around the
impurity. \cite{Lua92,Lua93}

The remainder of this paper is organized as follows: In Section \ref{aguado:theory}
we describe
the different clusters which have been used in order
to assert the influence of the embedding scheme and of the size of the
active cluster
on the results.
In Section \ref{aguado:results} we 
present and discuss the results of the calculations, and Section 
\ref{aguado:summary} summarizes
the main conclusions.

\section{Calculational strategy. Description of the different active clusters} 
\label{aguado:theory}

The {\em ab initio} Perturbed Ion model is a particular application of the
theory of electronic separability of Huzinaga and coworkers,
\cite{Huz71,Huz73} in which the basic building block is reduced to a single ion.
The PI model was first developed for perfect crystals. \cite{Lua90a}
Its application to the study of impurity centers
in ionic crystals has been described in refs. \onlinecite{Lua92,Lua93}, 
and we refer to those papers for a full account of the method. 
In this case, an active cluster containing the impurity is considered, and
the Hartree-Fock-Roothaan (HFR) equations \cite{Roo63} of each ion in the active
cluster are solved in the field of the other ions.
The Fock operator includes, apart from the usual intra-atomic terms, an
accurate, quantum-mechanical crystal potential
and a lattice projection operator which accounts for the energy contribution
due to the overlap between the wave functions of the ions. \cite{Fra92}
The atomic-like HFR solutions are used to describe the active cluster ions
in an iterative stepwise procedure.
Also, as we shall see later in more detail, the positions of some
ions in the active cluster are allowed to relax.
The wave functions of the lattice ions external to
the active cluster are taken from a PI calculation for the perfect crystal
and kept fixed during the embedded-cluster calculation.
Those wave functions are explicitely considered up to a distance $d$ from the 
center of the
active cluster such that the quantum contribution from the most distant frozen
shell to the effective cluster energy is less than 10$^{-6}$ hartree. In
practice, this means 25 shells for NaI:Tl$^+$ and 21 shells
for CsI:Tl$^+$.
Ions beyond $d$ contribute to the effective energy of the active cluster just
through the long-range Madelung interaction, so they are represented by point
charges.
At the end of the calculation, the ionic wave functions
are selfconsistent within the active cluster and consistent with the frozen
description of the rest of the lattice.  The intraatomic Coulomb correlation
correction, which is neglected at the Hartree-Fock level, is computed by
using the Coulomb-Hartree-Fock (CHF) model of Clementi. \cite{Cle65,Cha89}
An effective energy 
can be assigned to each ion that collects
all the contributions to the total crystal energy involving that ion. \cite{Lua90a}

In this work we have employed several active clusters of
increasing complexity and with different embedding schemes. Now we describe
each cluster (see fig. 1). First for NaI:Tl$^+$, that is, a
substitutional Tl$^+$ impurity in a NaI crystal with the rock-salt
structure:
\begin{itemize}
\item A) (TlI$_6)\ ^{5-}$:PC. This active cluster is formed by
the impurity cation Tl$^+$ and its 6 nearest neighbors, whose positions are
allowed to relax. The rest of the crystal is
simulated by using point charges (PC).
Notice that in this case, and also in cluster A$^*$ below, the environment is
simplified with respect to the statements at the beginning of this section, in
the sense that the ``frozen ions'' are just point charges.
\item B) (TlI$_6)\ ^{5-}$:NaI. The active cluster is the same as in A,
but now the environment is represented by ions whose ``frozen''
wave functions are obtained
from a PI calculation performed for the pure NaI crystal.
In the perfect crystal all the cations (or anions) are equivalent,
and there is a single nearest-neighbor Na--I distance. These characteristics
are lost around the substitutional impurity.
\item C) (TlI$_{14}$Na$_{18})\ ^{5+}$:NaI. The active cluster is formed by 33 ions
corresponding to the central impurity cation plus four coordination shells. The
rest of the crystal is represented by frozen ions with PI wave 
functions as in B. In the calculations
performed on this cluster only the positions of ions in the 
first shell around Tl$^+$ are allowed to
relax. The other three shells in the cluster
provide an interface between the first shell and the frozen environment. 
The ions in
this interface respond to the distortion induced by the impurity
by adapting selfconsistently their wave functions to the new potential, but
not their positions.
\item D) (TlI$_{92}$Na$_{86})\ ^{5-}$:NaI. This is the largest active cluster used
for this material. It has 179 ions which correspond to the central Tl$^+$
cation plus 12 coordination shells. Only the 33 ions of the four inner shells
relax their positions; the eight outer shells provide the interface 
between the inner ones and the
environment of frozen PI ions.
\end{itemize}
Next, for CsI:Tl$^+$, that is, a substitutional Tl$^+$ impurity in a CsI
crystal with the CsCl structure, we define four clusters in a similar way:
\begin{itemize}
\item A$^*$) (TlI$_8)\ ^{7-}$:PC. The active cluster is formed by the impurity plus
its eight nearest neighbors, and is embedded in an environment of point charges.
\item B$^*$) (TlI$_8)\ ^{7-}$:CsI. This cluster is 
similar to A$^*$, but the environment is now  
simulated by frozen ions with wave functions taken from a PI calculation for the
pure CsI crystal. 
\item C$^*$) (TlI$_{32}$Cs$_{32})\ ^+$:CsI. The active cluster has 65 ions and
is formed by the impurity plus six coordination shells. Only the positions
of the ions
in the first shell around Tl$^+$ are allowed to
relax. The environment is the same as in B$^*$.
\item D$^*$) (TlI$_{88}$Cs$_{92})\ ^{4+}$:CsI. The active cluster has 181 ions:
the impurity plus twelve coordination shells. The ions in the six innermost
shells can relax their positions, and
the environment is as in B$^*$.
\end{itemize}

The geometrical relaxations of the shells in the different cluster models
described above have been performed by allowing for independent
breathing displacements for each shell of ions, and minimizing the total
energy with respect to those displacements
until the effective
cluster energies are converged
up to 1 meV.
A downhill simplex algorithm \cite{Wil91} was
used.
For the ions we have used large STO basis sets: (5s4p)
for Na$^+$, \cite{Cle74} (11s9p5d) for I$^-$, \cite{Cle74} (11s9p5d) for Cs$^+$,
\cite{Mcl81} and (13s9p7d3f) for Tl$^+$. \cite{Mcl81}

\section{Results and discussion}
\label{aguado:results}

Before presenting the results for the lattice distortions around the Tl$^+$
impurity we test the consistency of the embedding method for the case of pure
(undoped) crystals. For this purpose we compare the results of two
sets of calculations for NaI. 
In the first one we study the clusters labelled A, B, C,
D in Section \ref{aguado:theory}, with the embedding scheme
indicated there for each case, but with a Na$^+$ cation instead of
the Tl$^+$ impurity. This is, of course, just the case of the pure NaI
crystal treated by the embedded cluster method. One can then compare the 
results with those of a standard PI calculation for the perfect crystal.
\cite{Lua90a} 
We can anticipate some differences between the two methods since in the usual
PI calculation for a perfect crystal
all the cations (or anions) are equivalent, while in the
embedded-cluster description of the same system the cation acting as a 
fictitious impurity and the other cations of the crystal are not described in
the same way. This systematic error is, in fact, what we want to remove in the
analysis of the distortions around the true, Tl$^+$, impurities.
If we call R$_1$ the distance between the Na$^+$ cation and
its first I$^-$ neighbors, Table I
gives R$_1$ for the four cluster models
A, B, C, D, together with the difference $\Delta R_1$ = [R$_1$(cluster) -
R$_1^{PI}$(crystal)] between the embedded-cluster result and that of the perfect
crystal at equilibrium and the relative deviation 
$\Delta R_1$/R$_1^{PI}$(crystal). The PI model predicts for the crystal an
equilibrium value R$_1^{PI}$(crystal) of 3.237 \AA, in very good agreement with the
experimental value R$_1^{exp}$ = 3.240 \AA. The first two cluster schemes, 
A and B, give large distortions $\Delta R_1$. 
In contrast, a contraction of R$_1$ smaller 
than 1 \% occurs for cluster models C and D. The conclusions for CsI are
similar. A good self-embedding is only achieved for clusters C and D (C$^*$
and D$^*$). Comparison of B and C shows the importance of a smooth interface
between the inner cluster core, where ions are allowed to move, and the
frozen environment around the cluster. Comparison of C and D establishes
the necessity of allowing for relaxation of the radii of several shells around
the impurity. The systematic error that the embedding-cluster scheme makes has
to be taken into account in order to interpret properly the distortions induced
by the Tl$^+$ impurity.

Now we present the results for the lattice distortions induced by the Tl$^+$
impurity in NaI and CsI. In this case we have calculated
\begin{equation}
\Delta R_i = R_i(NaI:Tl^+) - R_i(NaI:Na^+),
\label{aguado:distortion}
\end{equation}
where R$_i$ (i=1, 2, 3, ...) refer to the radii of the first, second, ...
shells around the impurity and both calculations have been performed in the
embedded cluster scheme. In this way the systematic errors of the cluster
method analyzed in Table I tend to cancel. 
As before, R$_1$ is the distance between the
central cluster cation, Tl$^+$ or Na$^+$, and the I$^-$ anions in its first
neighbor shell. For the reasons given above we only trust the results from
clusters C, D
(or C$^*$, D$^*$) and the results for R$_1$ and $\Delta R_1$ are given
in Table II, 
which also contains results for Tl$^+$ in CsI. Although we
discourage the use of models A and B, just for the purposes of comparison with
Berrondo {\em et al.} \cite{Ber96a} we have also calculated $\Delta R_1$ for
NaI:Tl$^+$ in model A and for CsI:Tl$^+$ in model A$^*$. 
Berrondo used a similar embedded-cluster scheme,
although his calculational method was different, based on the
linear combination of atomic orbitals (LCAO). Taking as reference
the experimental nearest-neighbor Na$^+$--I$^-$ distance in the bulk, Berrondo
obtained $\Delta R_1$ = 0.54 \AA, and taking the same reference we
obtain $\Delta R_1$ = 0.50 \AA. The good agreement between the two
calculations provides a check of our theoretical method. However this large
value of $\Delta R_1$ is partly due to the fact that the calculated
R$_1$(NaI:Tl$^+$) and the reference R$_1^{exp}$(pure crystal) are not fully
consistent with each other for the reasons discussed above. If we calculate
R$_1$(pure NaI) also by using model A (with Na$^+$ instead of Tl$^+$), that is,
like in eqn. \ref{aguado:distortion}, then a corrected ``smaller'' value $\Delta R_1$ = 0.20
\AA \ (or 5.6 \%) results.

Returning to models C and D we predict, using eq. \ref{aguado:distortion}
a lower expansion
of R$_1$. In fact, $\Delta R_1$ is equal to 0.039 \AA \ (1.2 \%
expansion) for model D.
However, the distortions are not restricted
to the first shell. The second and fourth shells also expand a little, while
the third shell suffers a contraction. 
The expansion of the first and second shells is consistent with
the fact that the ionic radius of Tl$^+$, equal to 1.40 \AA, is larger than the
ionic radius of Na$^+$ (0.95 \AA).
In summary we find that the
structural relaxation around the Tl$^+$ impurity in NaI is small, although
more complicated that usually assumed by previous workers. In fact our
calculations suggest that the relaxation could even go beyond the fourth shell.

Let us now turn to CsI:Tl$^+$. The calculation with model A$^*$ would give a
small expansion $\Delta R_1$ of the first shell if we take as
reference R$_1^{exp}$(pure crystal), and our calculation would then be in agreement
with that of Berrondo {\em et al.}. \cite{Ber96a} However the ionic radius of
Tl$^+$ is lower than that for Cs$^+$, so an expansion of R$_1$ is 
not to be expected.
By using, instead, eq. \ref{aguado:distortion} we obtained a corrected result
$\Delta R_1$ = $-0.062$ \AA, that is a contraction. Cluster
model C$^*$, in which only the first shell around Tl$^+$ is allowed to
distort, gives a small expansion of R$_1$ and finally model D$^*$ recovers
again the expected contraction. Beyond the first shell, $\Delta R_i$
oscillates ($\Delta R_2$ and $\Delta R_4$ are also negative, while
$\Delta R_3$, $\Delta R_5$ and $\Delta R_6$ are positive),
although the values of $\Delta R_i$ are very small, in fact smaller 
than in NaI:Tl$^+$. 

The conclusion from our calculations is that the
distortions around the impurity affect ions in several coordination shells
around the impurity, and that consideration of both the atomic and the
electronic relaxation of ions in those shells is required.

For each active cluster we have also 
calculated the absorption energy corresponding
to the intra-atomic transition from the singlet ground state
to the triplet excited state of the thallium ion
(6s$^2$($^1$S) $\rightarrow$ 6s$^1$6p$^1$($^3$P)). Following the Franck-Condon
principle, this absorption has to be calculated at the frozen 
ground state geometrical
configuration. Thus, to obtain the energy of the localized excitation, we
solve the HFR
equations \cite{Roo63} for that geometrical 
configuration of the active cluster, with
the Tl+ ion in the excited electronic state
6s$^1$6p$^1$($^3$P), and evaluate the
absorption energy as a difference of effective ionic energies:\cite{Mar92} 
\begin{equation}
E_{abs} = E_{eff}[Tl^+(^3P)] - E_{eff}[Tl^+(^1S)].
\label{aguado:absorption}
\end{equation}
To calculate the contribution of an electronic open shell to the intraatomic
energy of a LS electronic configuration, the Hartree-Fock-Roothaan formalism
requires the coupling constants of that specific LS term as input. 
For the description of the open shell ion we have used
the coupling constants of the $^3$P term as given in
ref. \onlinecite{Mal62} and the same basis set as for Tl+(6s2). 
The absorption spectra of these doped
crystals are more structured (there are several absorption bands); 
our calculated transitions should be identified 
with the band A of the absorption spectrum of NaI:Tl$^+$ (4.25 eV), \cite{Jac91}
and with the first absorption band of CsI:Tl$^+$ (4.27 eV). \cite{Nag95}
Both in NaI:Tl$^+$ and CsI:Tl$^+$,
the absorption energies we obtain become closer to the experiment as we
improve the description of the local geometry around the impurity, as Table
II shows.
In particular the absorption energies obtained with models D and D$^*$ are
remarkably accurate, and this can be ascribed to the accurate representation
achieved for the lattice distortion around Tl$^+$. Treatment of other
absorption bands would require to depart from some of the basic assumptions
of the PI model.

\section{Summary}
\label{aguado:summary}

We have reported a study of the local lattice distortions
induced by a Tl$^+$ impurity in NaI:Tl$^+$ and CsI:Tl$^+$ scintillators.
To that end, the {\em ab initio} Perturbed Ion (PI) model has been used.
Large active clusters, embedded in accurate quantum environments representing
the rest of the crystal, have been studied. The importance of performing
parallel calculations for the pure systems in order to suppress systematic
errors from the calculated local distortions has been stressed. The 
self-embedding performance of the method has been also analyzed with the result
that it is necessary to include in the active cluster some shells of ions with
frozen positions but able to adapt selfconsistently their
wave functions to the displacements of the inner shells. The local distortions
obtained extend beyond the first shell of neighbors in NaI:Tl$^+$.
Thus, the assumptions frequently employed in impurity-cluster calculations,
which consider the active space as formed by the central impurity plus its
first shell of neighboring ions  only, should be taken with some care. In CsI:Tl$^+$,
our results indicate that only the first shell suffers an appreciable
displacement. Calculated
absorption energies are in excellent agreement with the experimental
values, and that agreement improves as the description of the cluster
becomes more realistic.

Thus, the PI model is able to give a
realistic description of the geometrical distortion around a defect 
in an alkali halide crystal.
Of course, the advantages come
with some drawbacks: The simplifying assumptions present in the
actual version of the PI model make it nonsuitable
for the
study of excited states 
involving interatomic charge transfer from the impurity.
The method also has difficulties for accounting
for the crystal-field
splitting of open shell excited states.
These drawbacks concerning the treatment of excited states should not overlook
the essential message of this paper, that is, that the PI method combined with
an embedded cluster model allows to obtain accurately the distortion of the
lattice around impurity centers in ionic crystals.
A combined strategy, namely, PI calculations to obtain the
distortions, followed by a MO calculation on the optimized
geometry, can be a desirable option in the future.

$\;$

$\;$

{\bf Acknowledgements:} Work supported by DGES (PB95-0720-CD2-01) and Junta de
Castilla y Le\'on (VA63/96). 
A. Aguado is supported by a predoctoral fellowship
from Junta de Castilla y Le\'on. A. Ayuela acknowledges a Marie Curie fellowship
supported by the EU TMR program.

\newpage

{\bf Captions of Tables}

$\;$

{\bf Table I.}
Self-embedding test. Values of R$_1$, the radius of the first coordination
shell around Na$^+$ in NaI obtained from embedded-cluster calculations,
and their deviations $\Delta$R$_1$ from
the theoretical value obtained in a PI calculation for the perfect crystal.
Relative deviations $\Delta$R$_1$/R$_1^{PI}$(crystal) are given in brackets.
Note the good selfembedding achieved in cluster models C and D.

{\bf Table II.} 
Distortion induced by a Tl$^+$ impurity in NaI and CsI.
Distances R$_i$ (i=1, 2, ...) in \AA \
between the Tl$^+$ and its succesive coordination shells are given,
as well as the differences $\Delta$R$_i$
with respect to the values obtained in calculations for the perfect
crystal (also with the cluster model) and percent distortions. 
Last column gives absorption energies (see text) together with percent
deviations from the experimental value.

$\;$

$\;$

{\bf Captions of Figures}

$\;$

{\bf Figure 1.}
The different active clusters studied. Small spheres represent cations and large
spheres represent anions. (a) (TlI$_6)\ ^{5-}$;
(b) (TlI$_{14}$Na$_{18})\ ^{5+}$; (c) (TlI$_{92}$Na$_{86})\ ^{5-}$;
(d) (TlI$_8)\ ^{7-}$;
(e) (TlI$_{32}$Cs$_{32})\ ^+$; (f) (TlI$_{88}$Cs$_{92})\ ^{4+}$.
These clusters are embedded in the crystal as indicated in the text.

\newpage

\begin{table}
\label{aguado:selfembedding}
\begin{center}
\begin{tabular}{c c c c}

NaI:Na$^+$ & Cluster & R$_1$ (\AA) & $\Delta$R$_1$ (\AA) \\
\hline

 & A & 3.540 & + 0.303  \\

 & & & (+ 9.36 \%)  \\

 & B & 3.159 & - 0.078   \\

 & & & (- 2.41 \%)  \\

 & C & 3.213 & - 0.024  \\

 & & & (- 0.74 \%)   \\

 & D & 3.234 & - 0.003  \\

 & & & (- 0.09 \%)   \\
\hline

\end{tabular}
\end{center}
\caption{}
\end{table}

\onecolumn[\hsize\textwidth\columnwidth\hsize\csname
@onecolumnfalse\endcsname

\begin{table}
\label{aguado:geometry}
\begin{center}
\begin{tabular}{c c c c c c c c c c c}

NaI:Tl$^+$ & Cluster & R$_1$ & R$_2$ & R$_3$ & R$_4$ & $\Delta$R$_1$ & $\Delta$R$_2$ & $\Delta$R$_3$ & $\Delta$R$_4$ & Abs. (eV) \\
\hline

 & A & 3.738 & & & & + 0.198 & & & & 3.91 (-8.00 \%) \\

 & & & & & & (+5.59 \%) & & & & \\

 & C & 3.274 & & & & + 0.061 & & & & 4.19 (-1.41 \%) \\

 & & & & & & (+1.87 \%) & & & & \\

 & D & 3.273 & 4.692 & 5.510 & 6.612 & + 0.039 & + 0.075 & - 0.054 & + 0.043 & 4.23 (-0.47 \%) \\

 & & & & & & (+1.21 \%)  & (+1.62 \%) & (-0.97 \%) & (+0.64 \%) & \\
\end{tabular}

\begin{tabular}{c c c c c c c c c}

CsI:Tl$^+$ & Cluster & $\Delta$R$_1$ & $\Delta$R$_2$ & $\Delta$R$_3$ & $\Delta$R$_4$ & $\Delta$R$_5$ & $\Delta$R$_6$ & Abs. (eV) \\
\hline

 & A$^*$ & - 0.062 & & & & & & 4.17 (-2.34 \%) \\

 & & (- 1.62 \%) & & & & & & \\

 & C$^*$ & + 0.013 & & & & & & 4.21 (-1.40 \%) \\

 & & (+ 0.30 \%) & & & & & & \\

 & D$^*$ & - 0.047 & - 0.008 & + 0.003 & - 0.016 & + 0.006 & + 0.001 & 4.26 (- 0.23 \%) \\

 & & (- 1.14 \%) & (- 0.15 \%) & (+ 0.06 \%) & (- 0.21 \%) & (+ 0.09 \%) & (+ 0.01 \%) & \\

\end{tabular}
\end{center}
\caption{}
\end{table}

\newpage

\begin{figure}
\psfig{figure=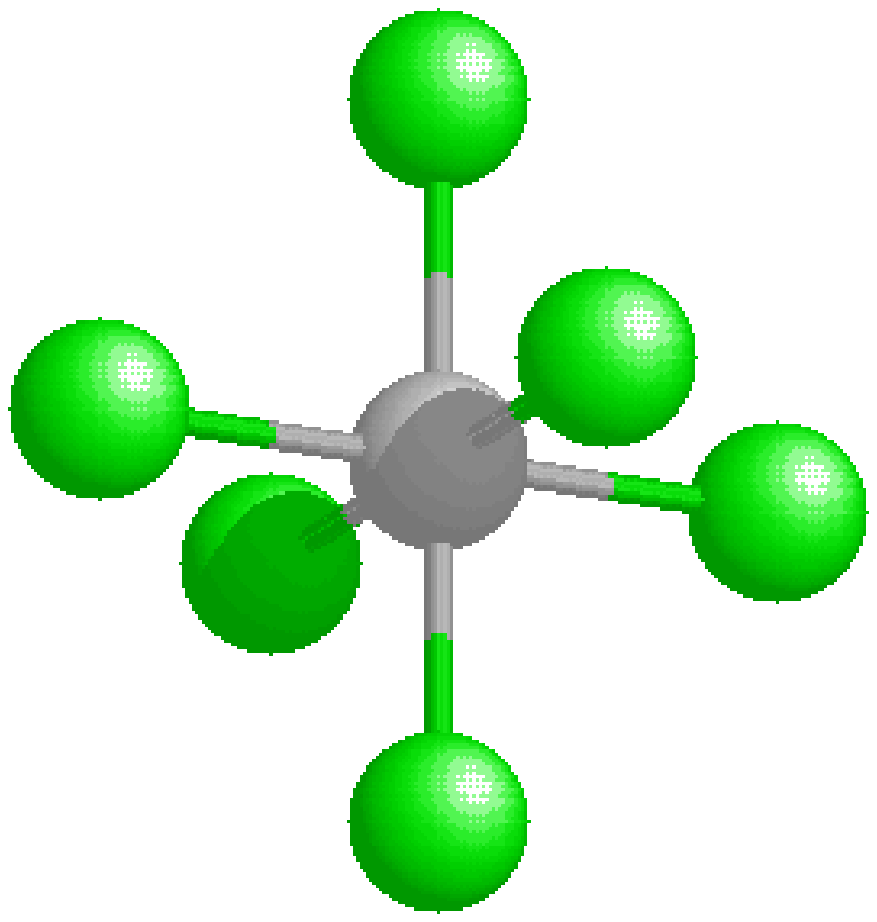}
\end{figure}

\begin{figure}
\psfig{figure=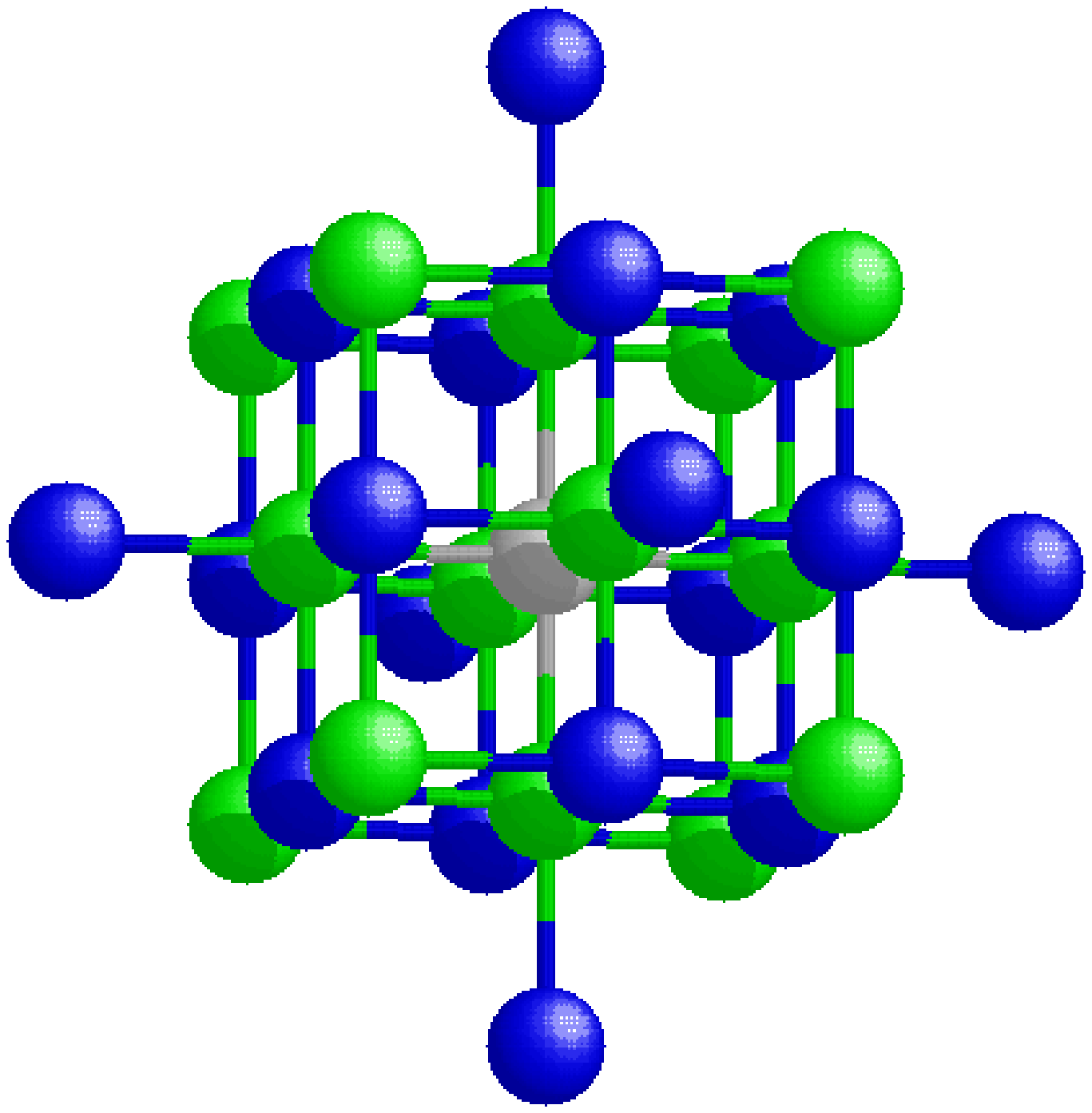}
\end{figure}

\begin{figure}
\psfig{figure=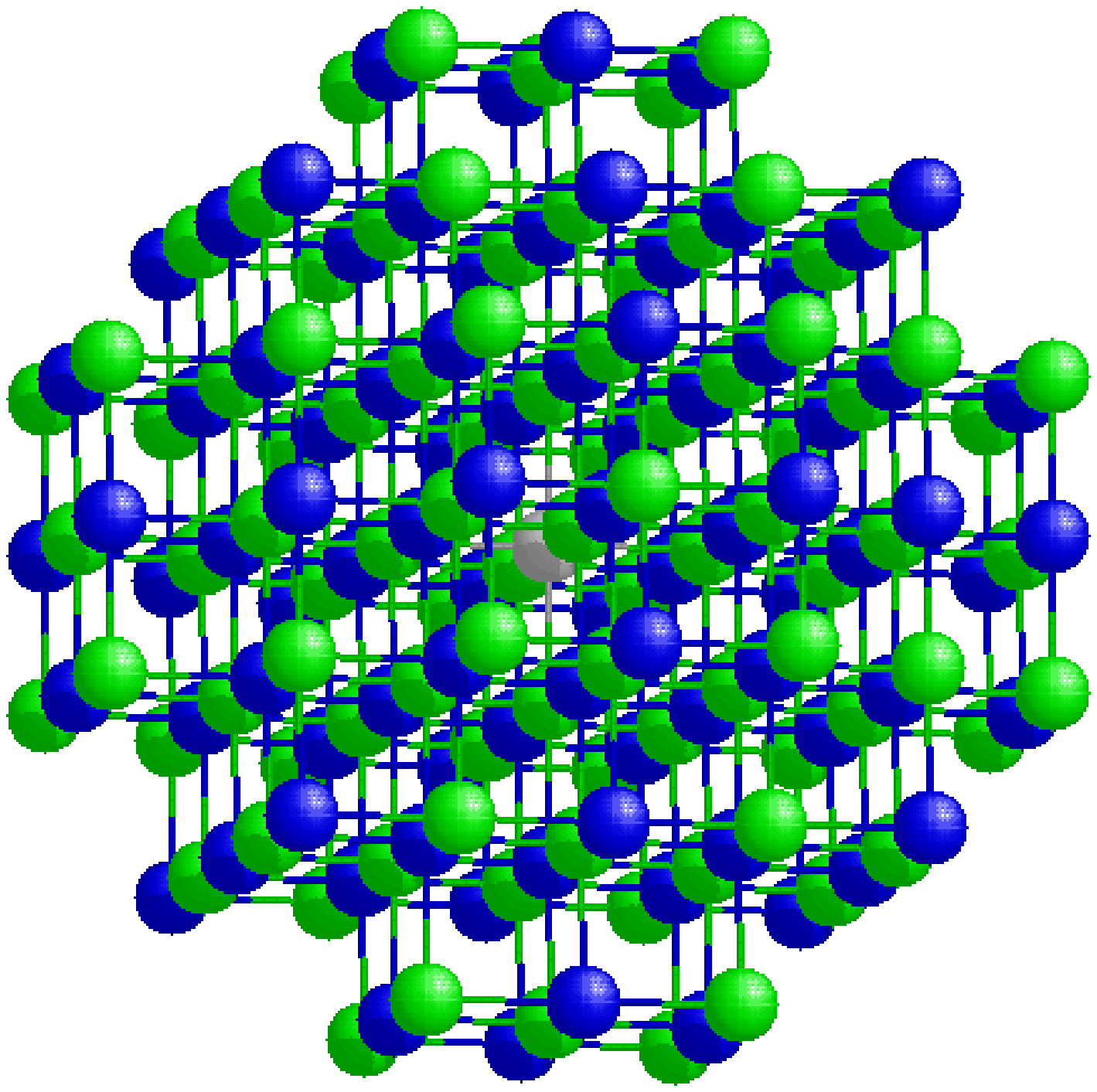}
\end{figure}

\begin{figure}
\psfig{figure=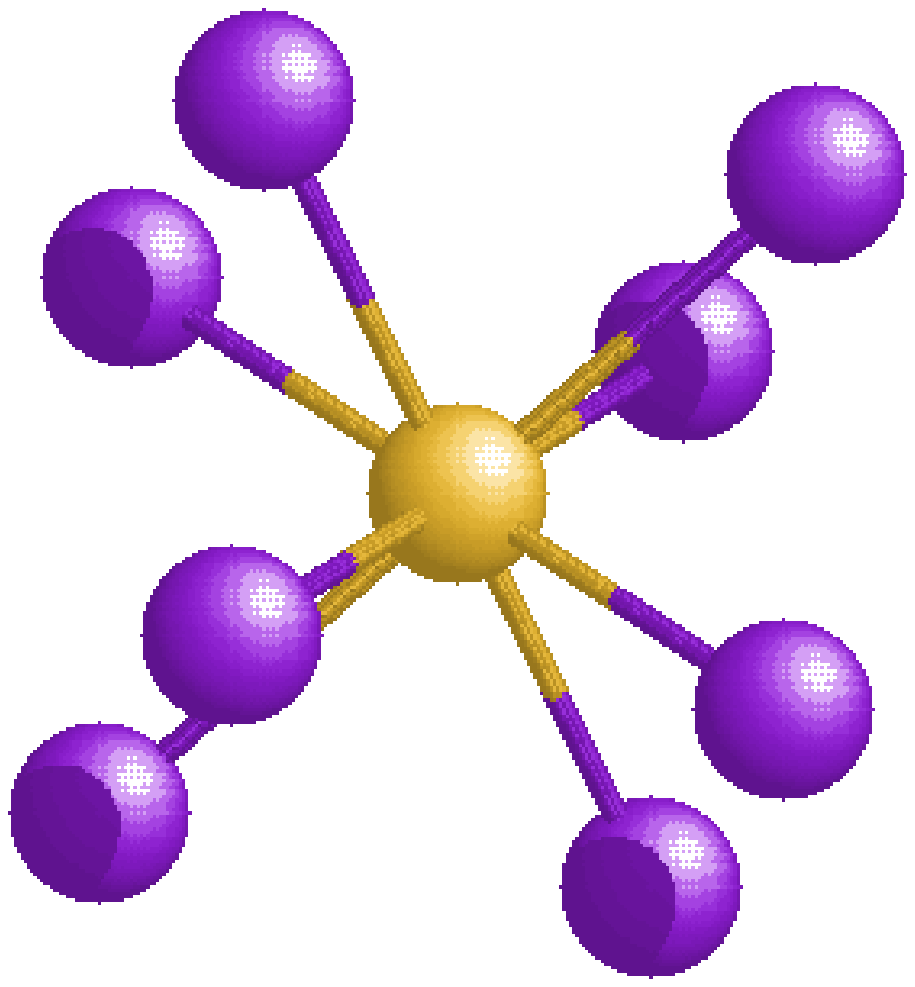}
\end{figure}

\begin{figure}
\psfig{figure=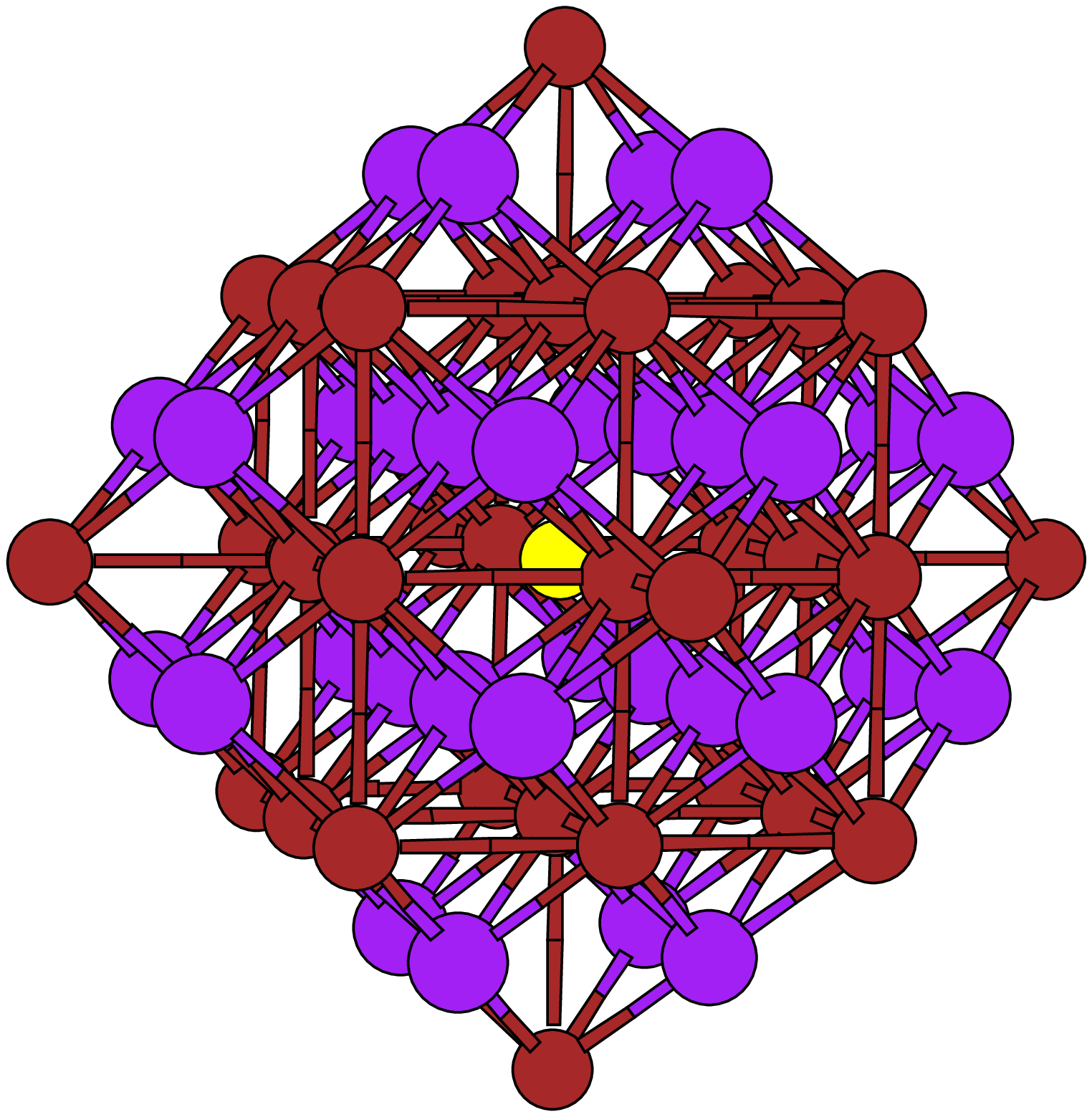}
\end{figure}

\begin{figure}
\psfig{figure=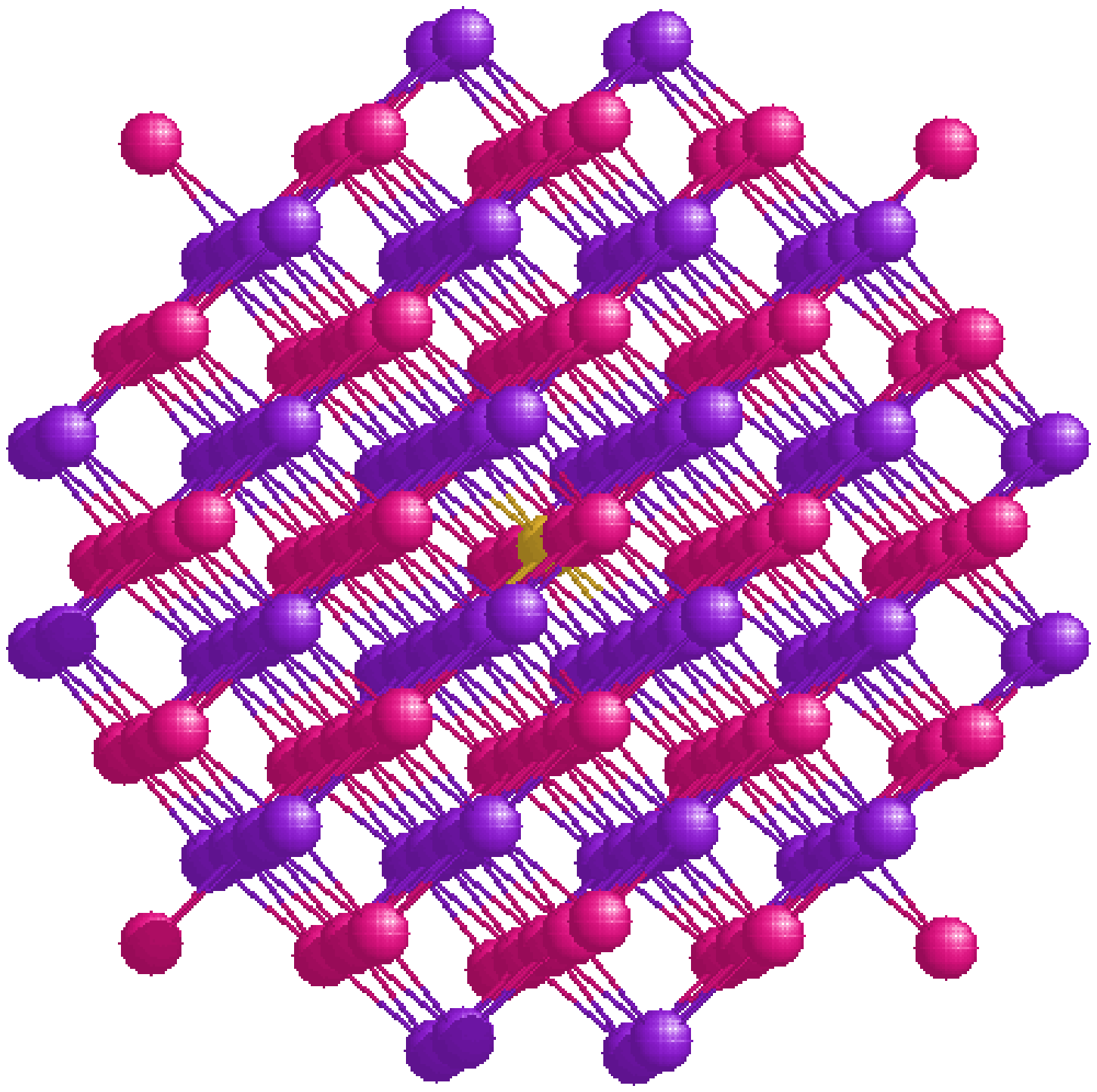}
\end{figure}

\end{document}